\documentclass{article}
\usepackage[utf8]{inputenc}
\usepackage{authblk}
\usepackage{setspace}
\usepackage[margin=1.25in]{geometry}
\usepackage{graphicx}
\graphicspath{ {./figures/} }
\usepackage{float}
\usepackage{subcaption}
\usepackage{amsmath}
\usepackage{graphicx}
\usepackage{multirow}
\newcounter{magicrownumbers}
\newcommand\rownumber{\stepcounter{magicrownumbers}\arabic{magicrownumbers}}

\usepackage{tabularray}
\usepackage{xcolor}

\usepackage{titlesec}
\usepackage{enumitem}

\titleformat{\paragraph}
{\normalfont\normalsize\bfseries}{\theparagraph}{1em}{}
\titlespacing*{\paragraph}
{0pt}{3.25ex plus 1ex minus .2ex}{1.5ex plus .2ex}

\usepackage[style=nejm, 
citestyle=numeric-comp,
sorting=none]{biblatex}
\title{Deployment and Analysis of Instance Segmentation Algorithm for In-field Grade Estimation of Sweetpotatoes}

\author[1*,4]{Hoang M. Nguyen}
\author[1,4]{Sydney Gyurek}
\author[2,4]{Russell Mierop}
\author[2,4]{Kenneth V. Pecota}
\author[3,4]{Kylie LaGamba}
\author[3,4]{Michael Boyette}
\author[2,4]{G. Craig Yencho}
\author[1,4]{Cranos M. Williams}
\author[1,4]{Michael W. Kudenov}

\affil[1]{Department of Electrical and Computer Engineering and NC Plant Sciences Initiative, \break North Carolina State University, Raleigh, U.S.}
\affil[2]{Department of Horticultural Science, \break North Carolina State University, Raleigh, U.S.}
\affil[3]{Department of Biological and Agricultural Engineering, \break North Carolina State University, Raleigh, U.S.}
\affil[4]{NC Plant Sciences Initiative, \break North Carolina State University, Raleigh, U.S.}
\affil[*]{Corresponding author. Email: hmnguye3@ncsu.edu}

\date{}

\begin{document}

\maketitle

\begin{abstract}
Shape estimation of sweetpotato (SP) storage roots is inherently challenging due to their varied size and shape characteristics. Even measuring “simple” metrics, such as length and width, requires significant time investments either directly in-field or afterward using automated graders. In this paper, we present the results of a model that can perform grading and provide yield estimates directly in the field quicker than manual measurements. Detectron2, a library consisting of deep-learning object detection algorithms, was used to implement Mask R-CNN, an instance segmentation model.  This model was deployed for in-field grade estimation of SPs and evaluated against an optical sorter.  Storage roots from various clones imaged with a cellphone during trials between 2019 and 2020, were used in the model’s training and validation to fine-tune a model to detect SPs.  Our results showed that the model could distinguish individual SPs in various environmental conditions including variations in lighting and soil characteristics.  RMSE for length, width, and weight, from the model compared to a commercial optical sorter, were 0.66 cm, 1.22 cm, and 74.73 g, respectively, while the RMSE of root counts per plot was 5.27 roots, with r$^2$ = 0.8. This phenotyping strategy has the potential enable rapid yield estimates in the field without the need for sophisticated and costly optical sorters and may be more readily deployed in environments with limited access to these kinds of resources or facilities. 
\end{abstract}

\newpage
\section{Introduction}

The sweetpotato, \emph{Ipomoea batatas} (L.) Lam. (2n=6x=90) is a globally important food crop.  Its storage roots are known for their starchy texture and sweet flavor, although this can greatly differ across varieties. Sweetpotato (SP) storage roots may be sold in the fresh market (e.g., food service, retail, international exports) or processed (e.g., canned, fried, frozen) \cite{SP_overview1, SP_overview_markets}. These storage roots (hereafter referred to as roots) are often qualified by visual traits such as shape, size, skin color, flesh color, and degree of defect. In different sectors of the SP market, buyers have different demands for the product's quality.  For example, a root's skin color and texture may be more relevant to consumers but less salient to canneries. Due to pigmentation from carotenoids and anthocyanins, the flesh color varies between varieties (white, yellow, orange, or purple). In regions where vitamin A deficiency is a significant public health concern, orange-fleshed varieties, rich in $\beta$-carotene, can supplement diets. Thus, for researchers and growers working with these varieties, flesh color can be an important characteristic \cite{SP_overview1, SP_overview_nutrition}. Within food production, quality of the crop, determined by the aforementioned traits, and consistency are highly valued.

To quantify quality and uniformity, roots are graded on several metrics. Length, diameter, and weight are size metrics commonly found in grade standards.  For example, the USDA specifies a U.S. No. 1 SP root as one with a diameter between 4.45 and 8.89 cm, a length between 7.62 and 22.86 cm, and a weight less than or equal to 567 g \cite{USDA_grades}.  The presence of disease, insect damage, physical damage, or other defects also contributes to a root's grade. Grade standards also typically have some shape specifications, as SP shapes play a large role in consumers' perception of a grade's quality. However, shape assessment with grading is highly subjective and difficult to standardize as consumer preferences are not concordant. The USDA's shape requirement for U.S. No. 1 roots is that they are "fairly well shaped" \cite{USDA_grades, SP_overview_markets2}. Modern industrial sorters and graders can quickly provide size estimates and identify external defects (discoloration, malformation, etc.); however, the infrastructure facility required for large-scale transportation, storage, and measurement of roots is costly. Logistically, the entire process is time-consuming and adds a risk of bruising the produce. For researchers, the need to transport and scan SPs at a single facility poses another problem: the possibility of cross-contamination when studying diseased roots.
A method to count and grade multiple roots in-situ would enable yield and quality estimates before packing, reducing some of the needs and costs associated with existing sorters. In addition, it would improve plant breeders ability to objectively phenotype large segregating populations of SP’s, which is  required for new variety development efforts.

Deep learning (DL) has been readily deployed for computer vision tasks. Object detection and instance segmentation using DL methods have been applied for yield estimation, automated harvesting, and disease detection of various horticultural products.  Ganesh \textit{et al.} demonstrates the detection of mangoes in trees using a combination of RGB and HSV images using Faster Region-based Convolutional Neural Network (R-CNN) \cite{deep_mangoes}. To classify between ripe and unripe strawberries and detect picking points, Yu \textit{et al.} utilized a combination of instance segmentation using Mask R-CNN and non-DL image processing methods \cite{deep_strawberry}.  In general, DL approaches are advantageous where variability in the subjects (shape, size, orientation) and scene (lighting, background) make it difficult to manually extract features to make decisions. DL models can be trained using labeled images as examples. Features relevant to identifying the subject are then found and used to make decisions in the inference step \cite{ag_survey, DL_Trad_Computer_Vision}.

 Different segmentation methods have been explored to estimate yield and sort potatoes. ElMasry \textit{et al.} performed segmentation by global color thresholding in the red channel to rapidly scan and classify potatoes by shape in a conveyor system. By using high contrast, black rollers and singulated potatoes, this segmentation approach was sufficient for the application \cite{potato_segmentation_conveyor}.  For segmentation with a low contrast background, such as a soil bed, more robust techniques are required. Lee \textit{et al.} developed an image processing pipeline consisting of filtering and morphological functions to isolate the potatoes from the soil, although it was necessary to space each potato so that none were touching one another \cite{potato_advance_image}.  More recently, using Mask R-CNN, Lee and Shin were able to isolate individual potatoes from a soil background, even with the potatoes directly in contact with each other \cite{maskrcnn_potato}.

This study investigated the efficacy of a deep learning (DL) model for instance segmentation of sweetpotato storage roots in which a cellphone-based field sampling protocol was implemented using images of SPs prior to grading in the warehouse, offering preliminary grade distributions prior to packing. We leverage Facebook AI Research (FAIR) group's Mask R-CNN \cite{MaskRCNN} - as implemented in part of the Detectron2 library \cite{Detectron2}. Individual SP shape properties are then extracted from the collected RGB cellphone camera imagery. A model was then created to convert a SP's area into a weight estimate. This method achieves estimates in common metrics (length, diameter, weight) comparable to those from SP's scanned by an optical grader.
In section \ref{preliminary}, we describe a preliminary experiment investigating the correlation between 2D root images and root weight. In section \ref{preliminary}, we discuss our procedure towards developing this model. In section \ref{Model Training}, we discuss the materials and methods used to train our model.  In section \ref{Model Validation}, we detail the methods used to validate our model. This consisted of plot-level and individual one-to-one comparisons between our model and an optical sorter.

\section{Materials and Methods}
\subsection{SP Weight Estimation From 2D View}
\label{preliminary}
To quantify the impact that a root's orientation has in the cellphone model, simulations were conducted to investigate the relationship between the distributions of possible 2D views of the SP and its volume. 3D models of sweetpotatoes were created and used in a physics-based Monte Carlo simulation in Blender, creating a wide range of 2D projections that could be analyzed against the known volume and weight \cite{blender}. A high degree of correlation would support using 2D instance segmentation for predicting 3D metrics.

\subsubsection{3D Reconstruction of SP}
\begin{figure}[h]
    \centering
    \begin{subfigure}{0.32\textwidth}
        \includegraphics[width=\textwidth]{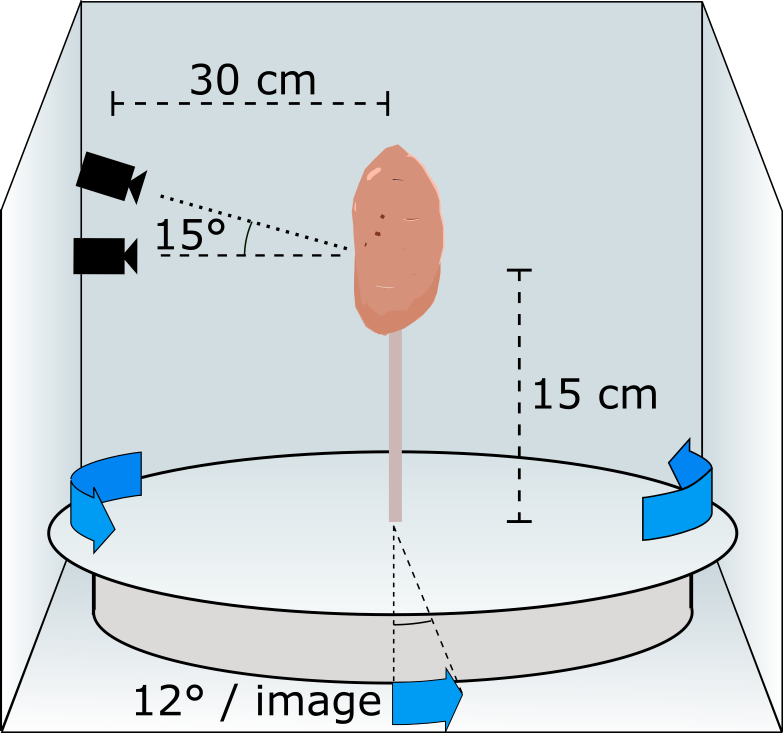}
        \caption{\label{fig:scan_a}}
    \end{subfigure}
    \begin{subfigure}{0.32\textwidth}
        \includegraphics[width=\textwidth]{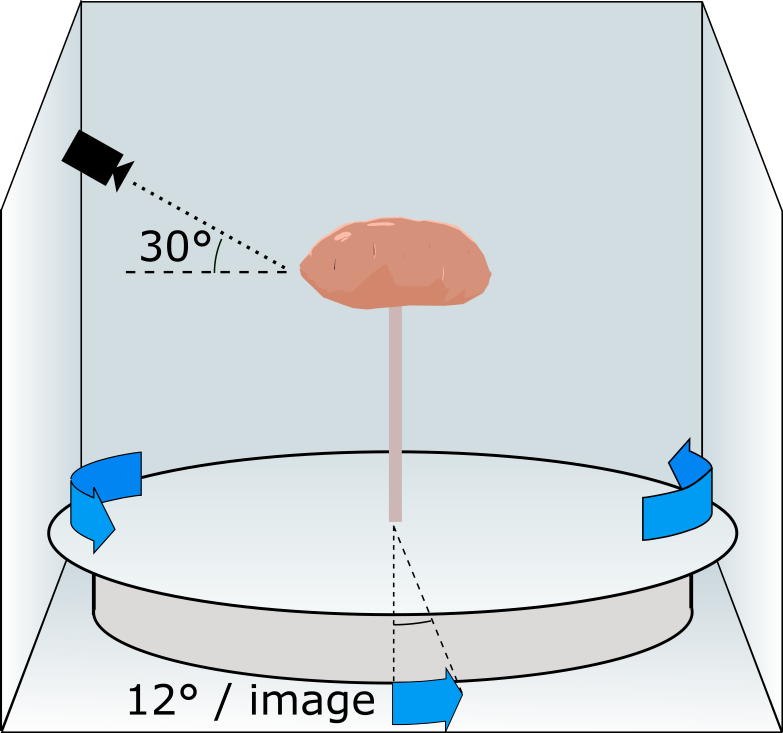}
        \caption{\label{fig:scan_b}}
    \end{subfigure}
    \begin{subfigure}{0.32\textwidth}
        \includegraphics[width=\textwidth]{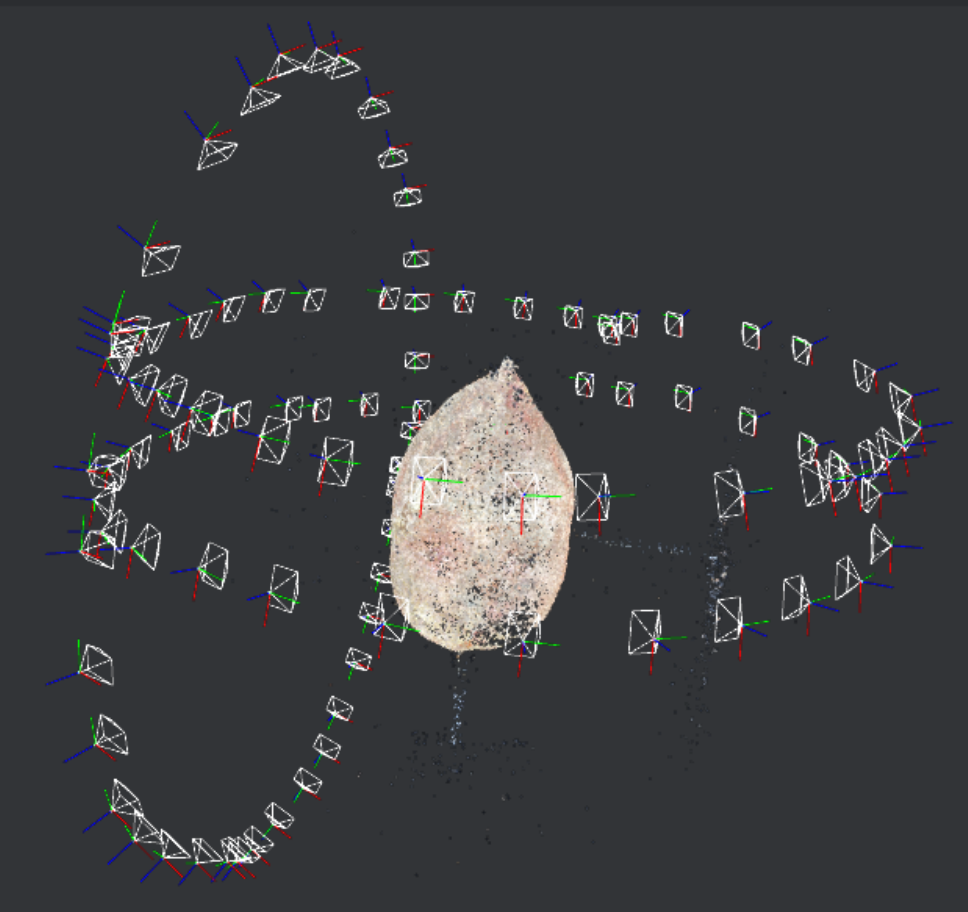}
        \caption{\label{fig:scan_c}}
    \end{subfigure}

\caption{For photogrammetry of the storage roots, each root was enclosed on three sides with a black backdrop.  (\subref{fig:scan_a}) The SP was first oriented parallel to its long (major) axis and mounted to a turntable. Two sets of images, one perpendicular to the axis and the other 15 deg from the horizontal, were captured with the camera. (\subref{fig:scan_b}) The turntable was rotated from 0 to 360 degs in about 12 deg increments between images. The root was then oriented  parallel to its short (minor) axis and the same imaging process was performed.  (\subref{fig:scan_c}) From the 3D viewer, the reconstructed model with the computed camera positions and angles from Meshroom.}
\label{fig:3D_scan}

\end{figure}

3D models of 18 jumbo SPs were created for a Monte Carlo simulation.  Using Meshroom, a software package based on the photogrammetry framework AliceVision \cite{meshroom}, multi-angle SP images were used to create the models. The imaging process is illustrated in Figure \ref{fig:3D_scan}. Using a 12-megapixel camera, about 80 to 100 images were captured for each SP. To minimize reconstruction time, a maximum viewing angle of 30$^{\circ}$ was set for the depth map and depth map filtering stages.  
 The measurements resulted in high-resolution 3D models, with meshes containing approximately 10$^5$ faces. Holes in the meshes at the skewered points were filled by extrapolation in Blender. Each model was then scaled according to the measured weight of the SP, assuming a constant density of an SP to be 1 g/cm$^3$ \cite{density}.

\subsubsection{3D-to-2D Monte Carlo Simulations}
\label{Monte Carlo}

A Python script was written to run 3D physics-based simulations using Blender \cite{blender} as Blender provided capabilities for mesh editing directly from Meshroom's \cite{meshroom} reconstruction, rigid body physics simulations, and built-in Python scripting. Simulations of SPs falling onto a flat surface and rollers were modeled to examine the effect of the roots' orientation on estimates of 3D metrics.

\begin{figure}[h]
    \centering
    \begin{subfigure}{0.32\textwidth}
        \includegraphics[width=\textwidth]{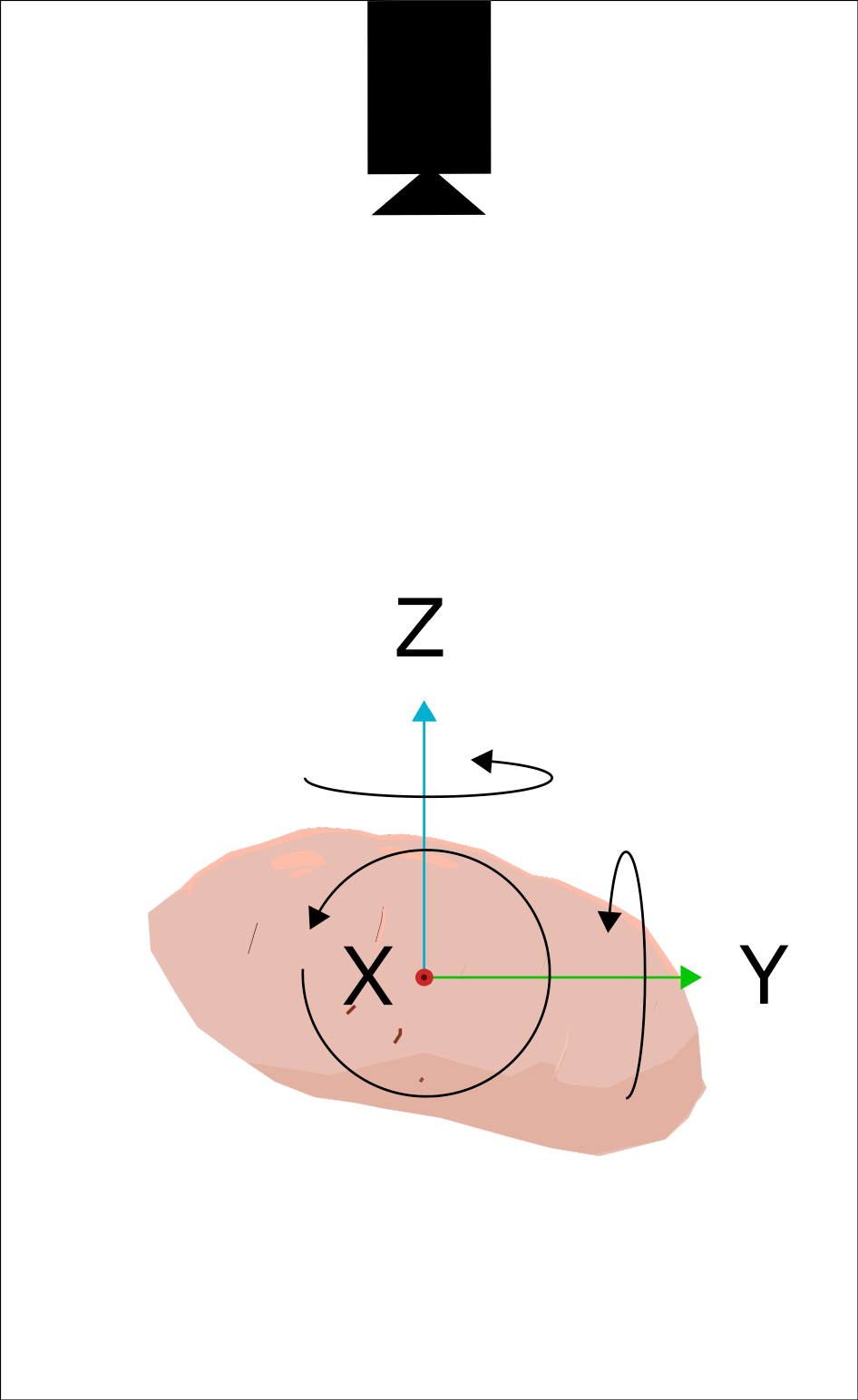}
        \caption{\label{fig:dof_1}}
    \end{subfigure}
    \begin{subfigure}{0.32\textwidth}
        \includegraphics[width=\textwidth]{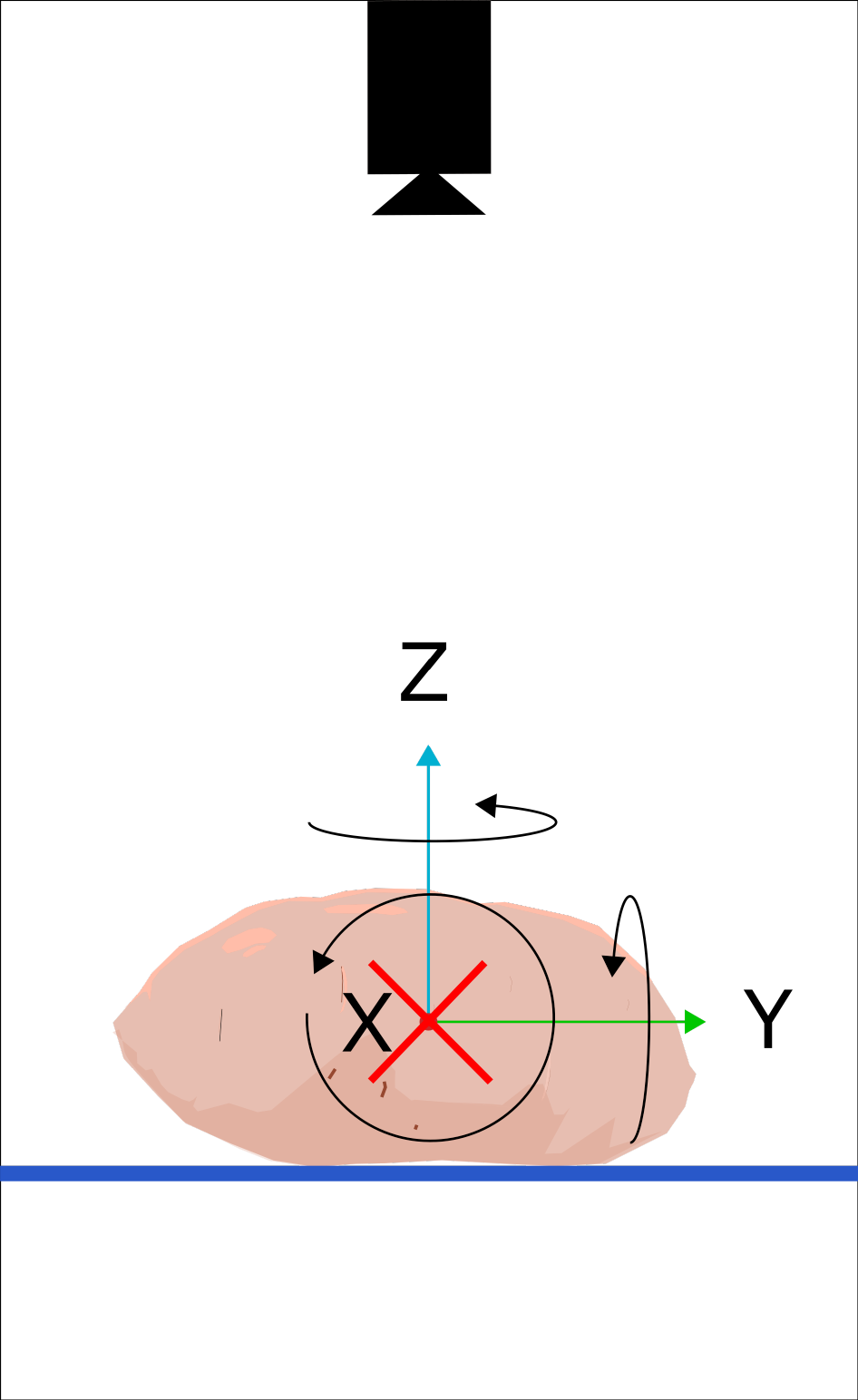}
        \caption{\label{fig:dof_2}}
    \end{subfigure}
    \begin{subfigure}{0.32\textwidth}
        \includegraphics[width=\textwidth]{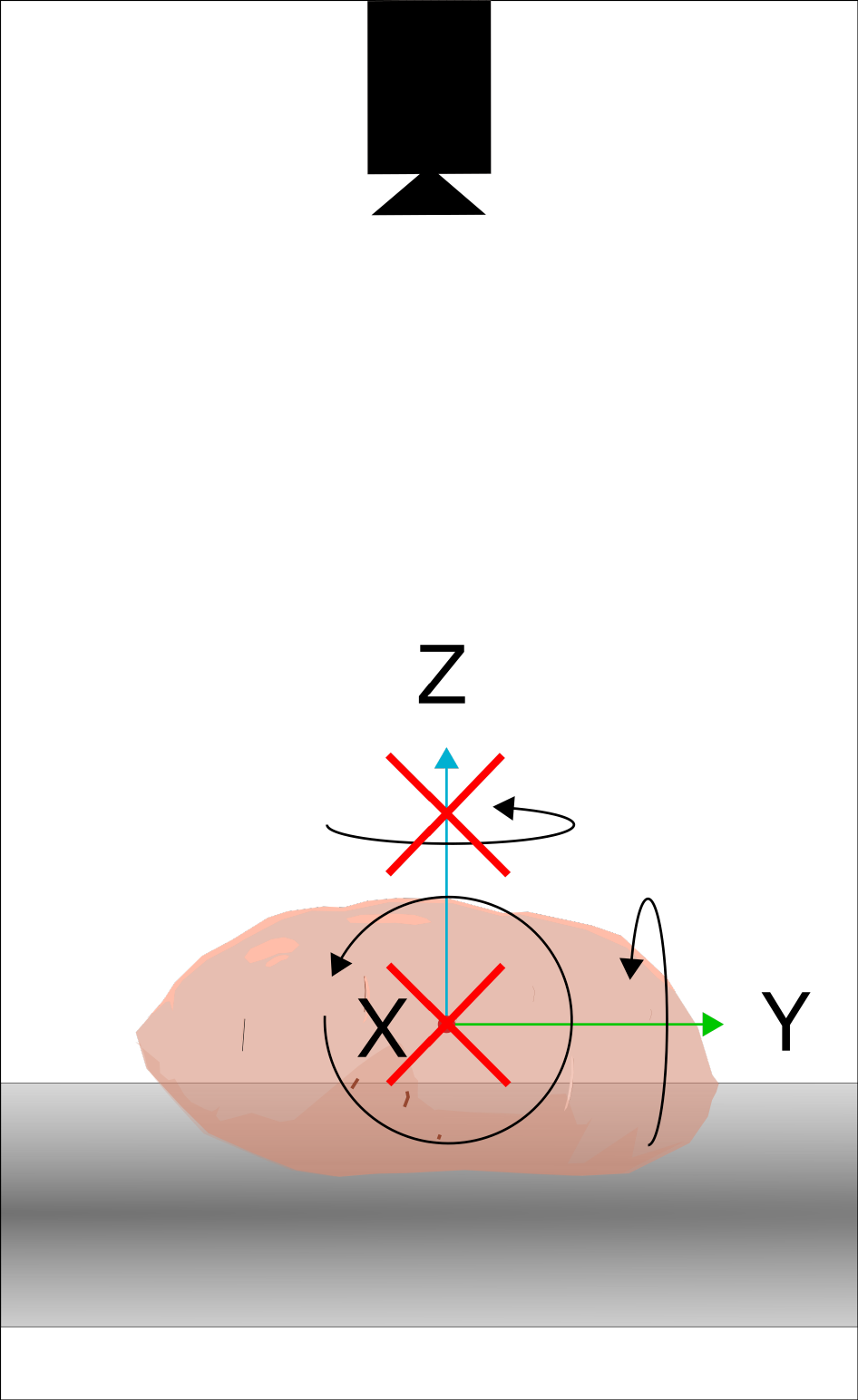}
        \caption{\label{fig:dof_3}}
    \end{subfigure}

    \caption{ The following constraints were considered in this simulation: (\subref{fig:dof_1}) free space; (\subref{fig:dof_2}) plane; and (\subref{fig:dof_3}) rollers.  A camera view of the SP from the positive z-axis was simulated.}
    \label{fig:dof}
    
\end{figure}

As illustrated in Figure \ref{fig:dof}(\subref{fig:dof_1}), initial simulations were performed considering the domain of all possible 2D profiles from the 3D SP in free space. As SPs are generally asymmetric, it was hypothesized that a poor correlation between the cross-sectional area and volume would be obtained. While it is unlikely that a root could be orientated arbitrarily, this can provide a baseline against which to perform comparisons. Conversely, in Figure \ref{fig:dof}(\subref{fig:dof_2}), the SP was constrained to a flat plane. Under these conditions, it was hypothesized that the correlation between cross-sectional area and volume should improve, as rotation about the \textit{y}-axis was constrained. It should be noted that this geometry is also most similar to the intended use case of imaging SPs on the ground using the cellphone camera. Finally, in Figure \ref{fig:dof}(\subref{fig:dof_3}), the plane was changed to a series of simulated rollers. In this case, the rotation of the SP about the \textit{y}- and \textit{z}-axes were constrained. This simulation was conducted to quantify performance when SPs are imaged on eliminator tables, which are common in many commercial SP processing and packing operations for hand-picking culls and automatically removing small debris. In all cases, the final position of the camera's view angle was constrained such that it was parallel to the \textit{z}-axis.

\begin{figure}
    \centering
    \begin{subfigure}{0.3529\textwidth}
    \centering
        \includegraphics[width=0.7\textwidth]{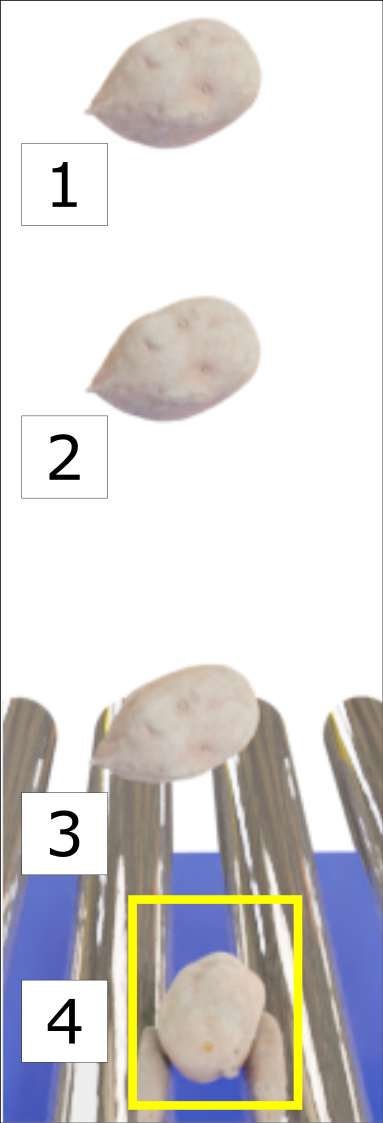}
        \caption{\label{fig:bsim_a}}
    \end{subfigure}
    \begin{subfigure}{0.637\textwidth}
    \centering
        \includegraphics[width=0.7\textwidth]{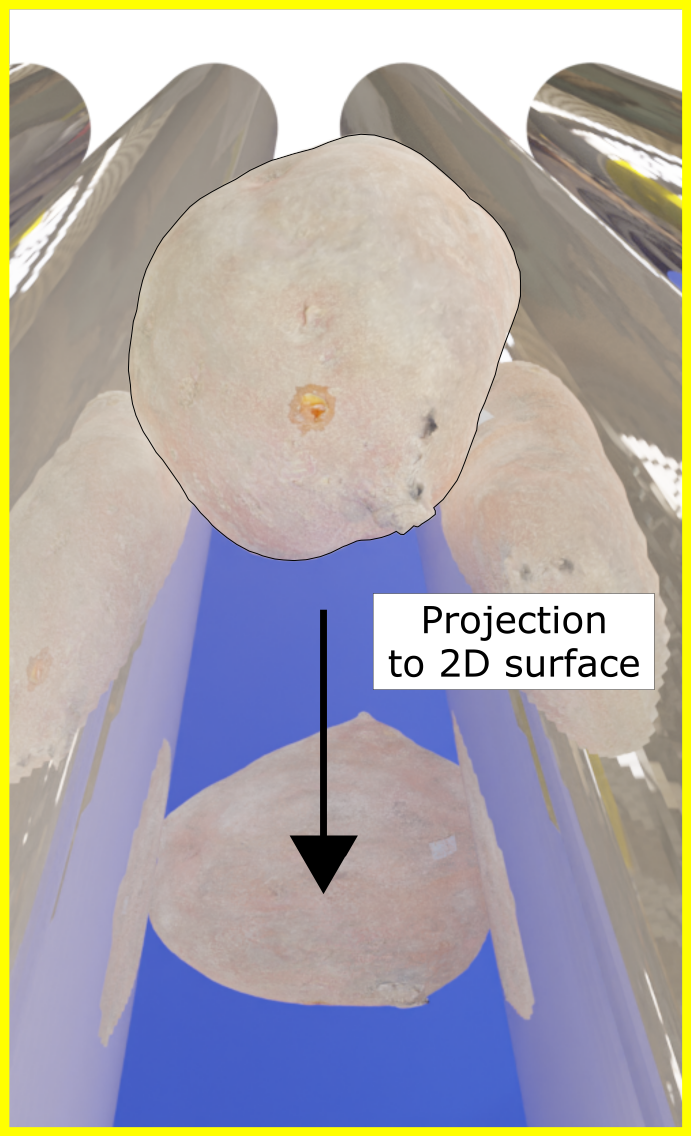}
        \caption{\label{fig:bsim_b}}
    \end{subfigure}

    \caption{Frames from a single simulation are shown in (\subref{fig:bsim_a}). Labels 1-4 illustrate the SP at frames 0, 30, 60, and 100. In  
    (\subref{fig:bsim_b}), the final frame (label 4) is projected normal to the blue plane, simulating imaging the root from above with a telecentric camera.  Note that the projection is not occluded by the rollers.
    \label{fig:bsim}
    }
\end{figure}

For the simulation setup, the model mesh's total number of faces were reduced by applying the "Decimate" modifier with a 0.01 ratio. To create 2D projections of the model, the "Shrinkwrap" modifier was set but not applied, targeting a plane, where the wrap method was set to "Target Normal Project". By not applying the modifier, this enabled changes in the mesh's orientation to update the 2D projection. A uniform random rotation angle was then generated using Shoemake's algorithm \cite{Shoemake_rotation}. For independent random variables $A, B, C$ uniformly distributed from 0 to 1 (exclusive), a uniform random rotation in 3D can be generated as a quaternion: 

\begin{equation} \label{eq:1}
    \mathbf{Q} = W + \mathbf{i}X + \mathbf{j}Y + \mathbf{k}Z,
\end{equation}

where the basic quaternions are defined as 
\begin{equation}
\label{eq:6}
    \mathbf{i^2} = \mathbf{j^2} = \mathbf{k^2} = \mathbf{ijk} = -1.
\end{equation}
The parameters are defined as

\begin{equation}
 W = \sin{(2\pi A)}\sqrt{1 - C}, 
 \end{equation}
\begin{equation}
 X = \cos{(2\pi A)}\sqrt{1 - C},
 \end{equation}
 \begin{equation}
 Y = \sin{(2\pi B)}\sqrt{C}, \mathrm{ and}
 \end{equation}
 \begin{equation}
 Z = \cos{(2\pi B)}\sqrt{C}.
 \end{equation}

For the free space simulations, after the SP mesh was randomly oriented, its projected area to the plane below was recorded without a rigid body simulation. Meanwhile, Figure \ref{fig:bsim} shows the process that was used for constrained simulations versus time. For the plane and roller constraints, the sweetpotato mesh was randomly positioned and dropped at 0.5 m above the plane or rollers and the rigid body physics was simulated for a total of 100 frames. The last frame was again used to record the SP's projected area, as based on the appropriate camera viewing geometry (e.g., parallel to the \textit{z}-axis). 

The SP's volume ($V_{SP}$) was then estimated from the projected area ($A_{proj}$) using two volumetric models: (1) ellipsoid model and (2) square-cube model. Assuming an ellipsoidal volume with circular cross-sections perpendicular to its major axis, for semi-axes lengths, \textit{a}, \textit{b}, and \textit{c}, the equation for the ellipsoid's volume ($V$) is

\begin{equation}
 V_{SP} \approx  V_{ellipsoid} = \frac{4\pi}{3}abc.
\end{equation}
 Assuming the projected area is an ellipse such that
 \begin{equation}
 A_{proj} \approx  A_{ellipse} = {\pi}ab,
 \end{equation}
 
\noindent if \textit{a} is the semi-major axis of the ellipsoid, the estimated volume is

\begin{equation}
 V_{SP} \approx \frac{4}{3}A_{proj}c,
 \label{eq:ellipse_volume}
 \end{equation}

Assuming a circular cross-section ($b = c$), Eq. \ref{eq:ellipse_volume} can be used to estimate the 3D volume.  This model is most accurate when the projected area is closest to the major ellipse.  However, due to the variability of orientations, this assumption might not be accurate.  A symmetric model may account for this. The square-cube model is given by

\begin{equation}
 V_{SP} \approx (A_{proj})^{\frac{3}{2}}
 \label{eq:volume}
 \end{equation}

It should be noted that these models were chosen, as opposed to a data-driven model (either regression or machine learning) or a complex shape model, to confine the study's scope. This expression produces certain quantifiable biases that could be resolved using alternative methods to convert $A_{proj}$ to $V$, the comparison and analysis of which are the subject of a future study.

\subsection{Mask R-CNN for SP Instance Segmentation}
\subsubsection{Model Training and Setup}
\label{Model Training}
\begin{figure}[H]
    \centering
    \includegraphics[width=0.5\textwidth]{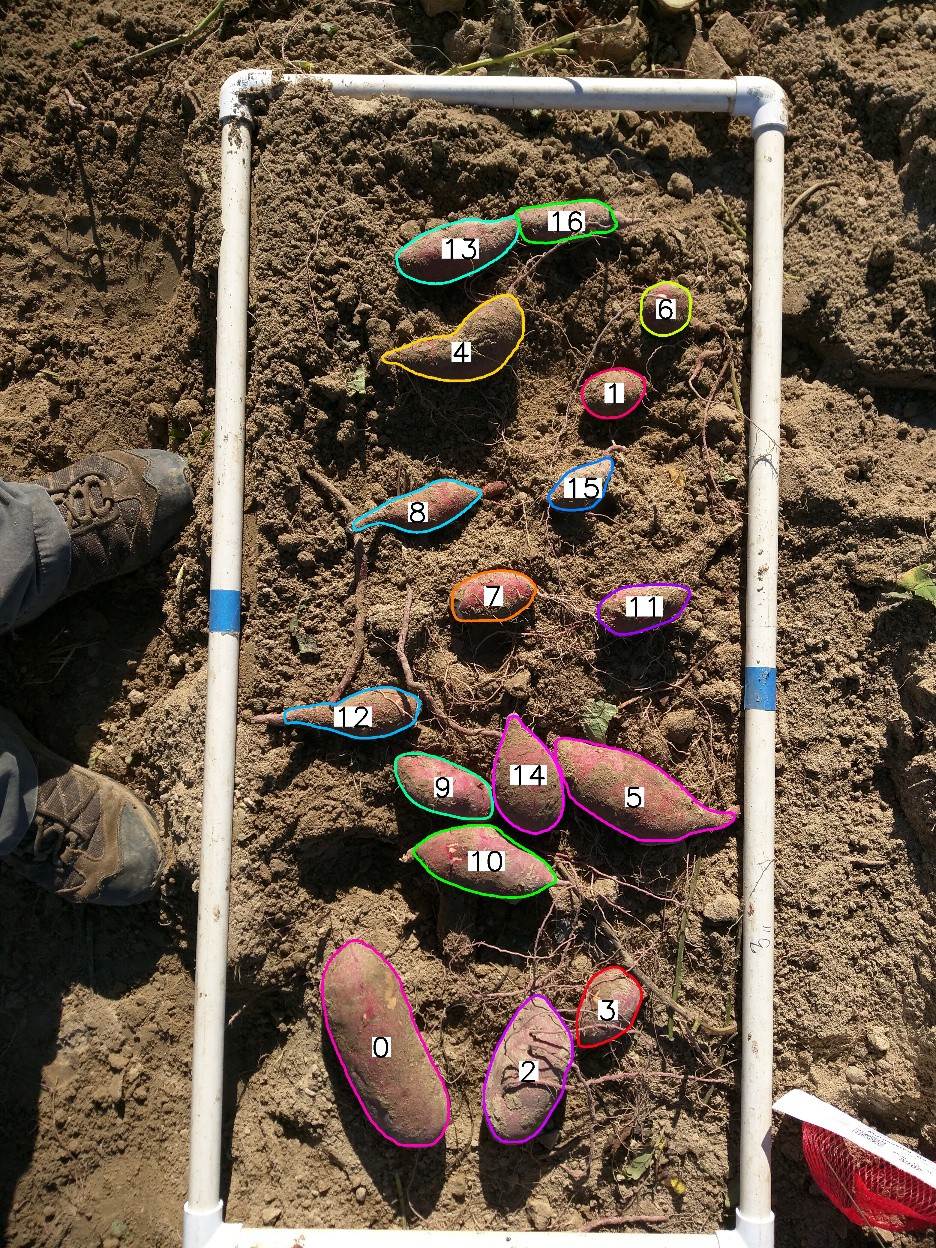}
    \caption{An example inference from Mask R-CNN used in the cellphone protocol.}
    \label{fig:mask_rcnn}
\end{figure}

Transfer learning of a Mask R-CNN model to detect SPs involved collecting a series of root images at various trial sites, outlining sweetpotatoes in the collected imagery, defining training parameters, and training the model using pre-trained weights. To utilize the masks, a spatial calibration was used to identify the number of cm/pixel. An example output from the Mask R-CNN model that identified individual SPs is shown in Figure \ref{fig:mask_rcnn}.  

\paragraph{Trial Sites} The SPs used in fine-tuning the model were obtained from trials conducted by Sweetpotato Breeding and Genetics program at NC State University during 2019 and 2020. SPs were harvested and imaged in September and October of each trial year. Locations included Cunningham Research Station in Kinston, NC (Lenoir County) and NCDA\&CS's Horticultural Crops Research Station (HCRS) in Clinton, NC (Sampson County).

\begin{figure}[!h]
\centering

\valign{#\cr
  \hsize=0.48\textwidth
  \subfloat[]{\includegraphics[width=\hsize]{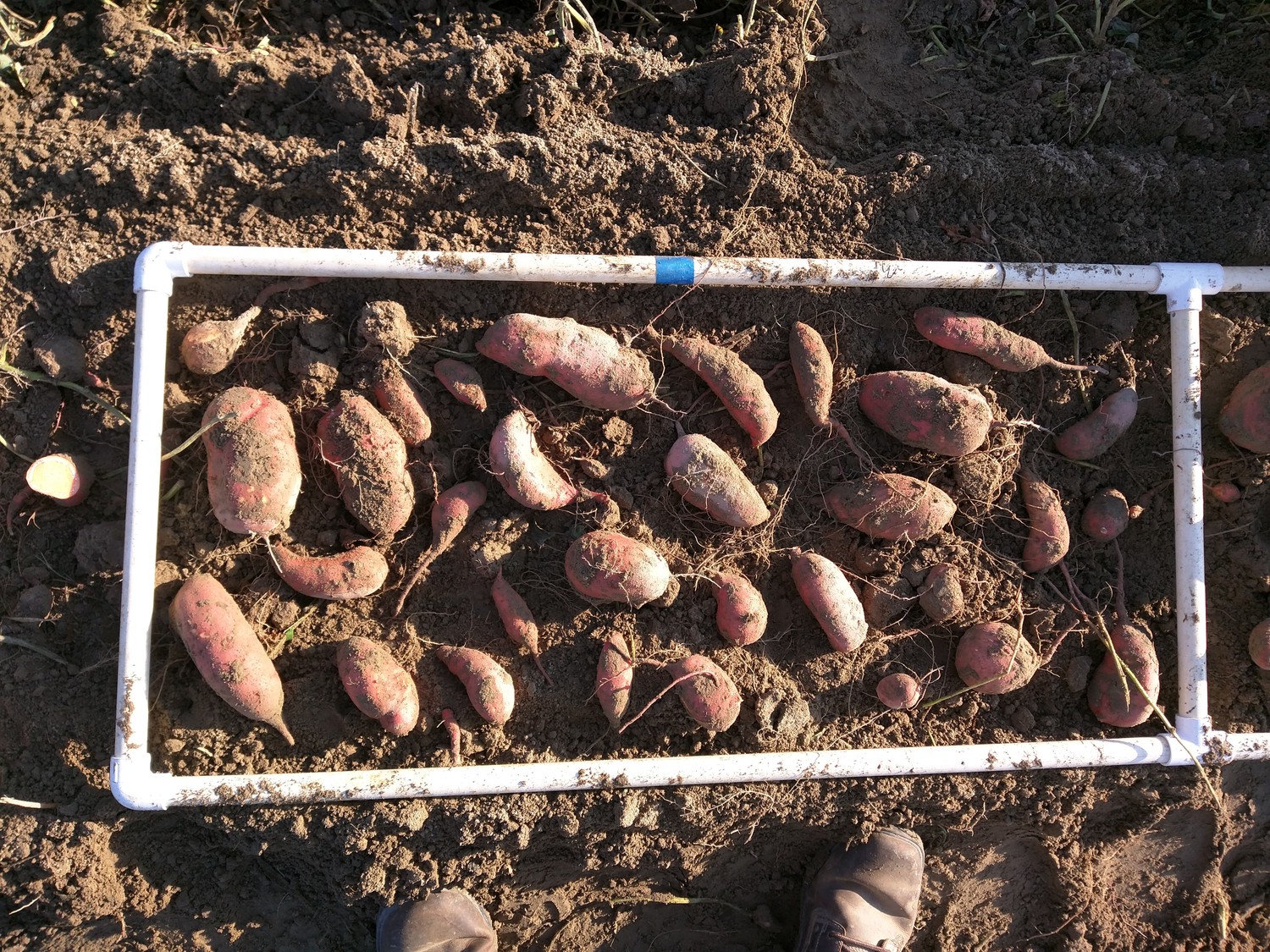}\label{fig:plot_1}}\vfill
  
  \subfloat[]{\includegraphics[width=\hsize]{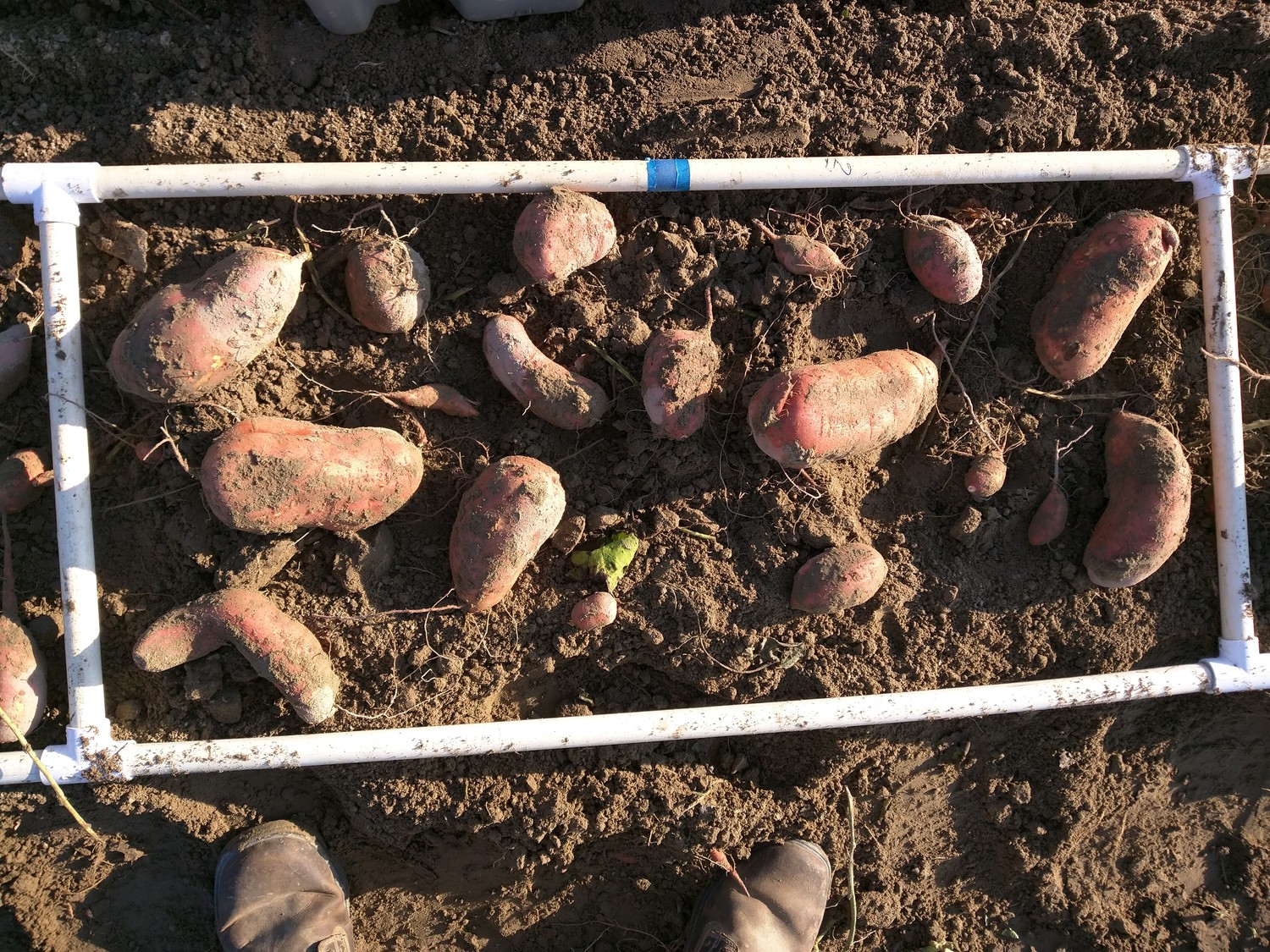}\label{fig:plot_2}}\vfill
  
  \subfloat[]{\includegraphics[width=\hsize]{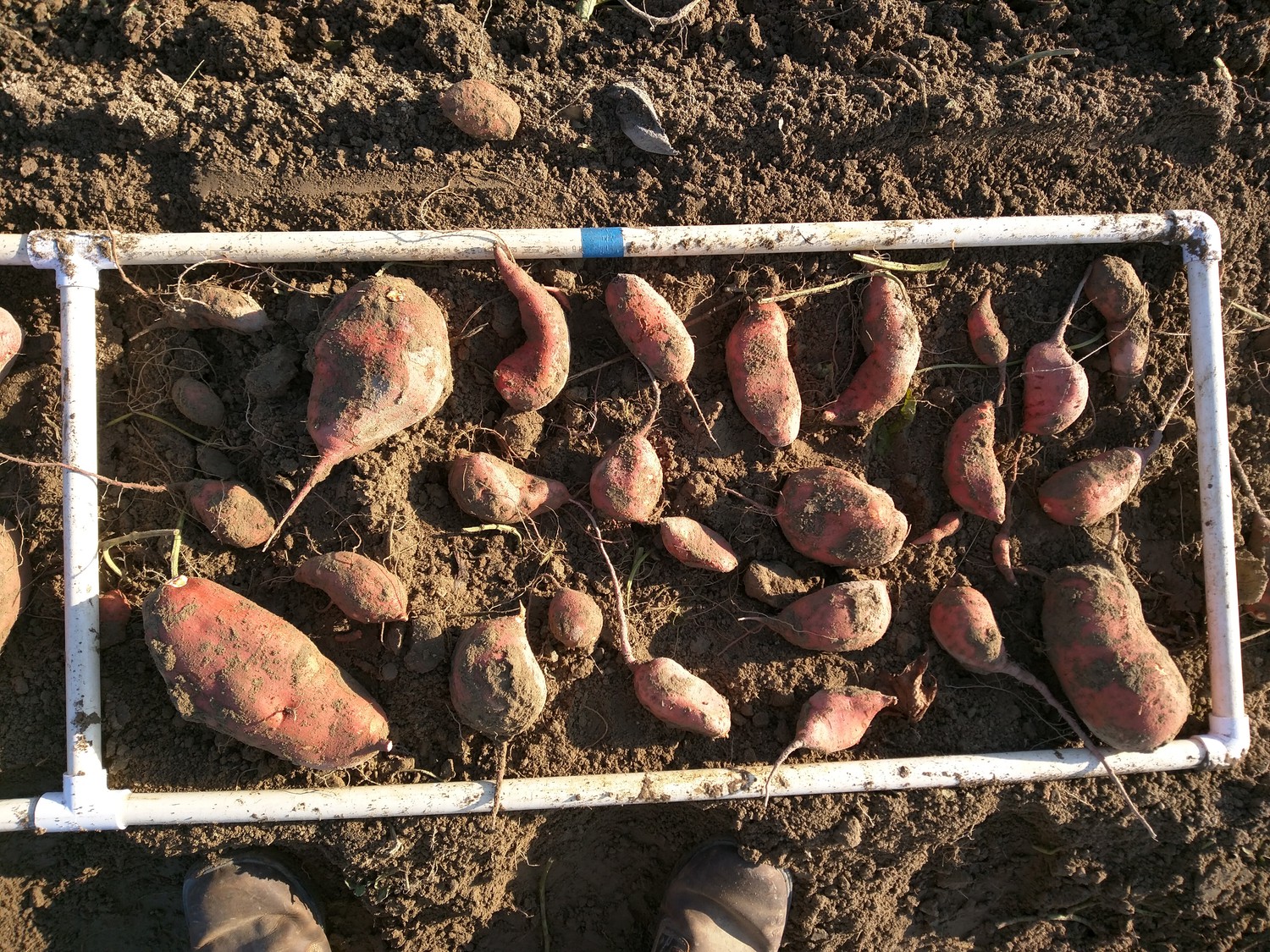}\label{fig:plot_3}}
  
  \cr\noalign{\hfill}
  \hsize=0.48\textwidth
  \subfloat[]{\scalebox{1}[1.05]{\includegraphics[width=\hsize, height = 11.1cm]{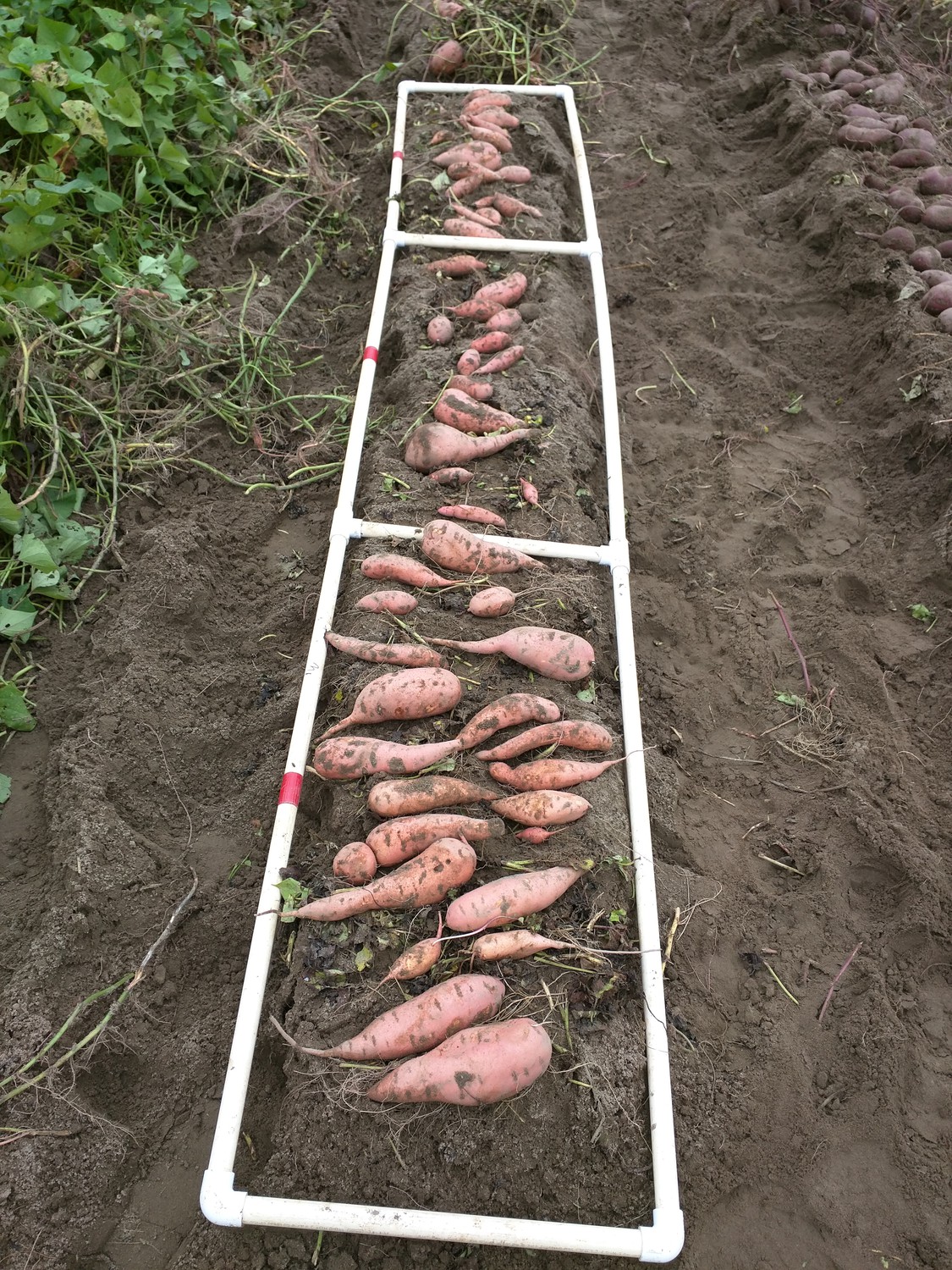}}\label{fig:offaxis}}\vfill
  \subfloat[]{\includegraphics[width=\hsize, height = 5.5cm ]{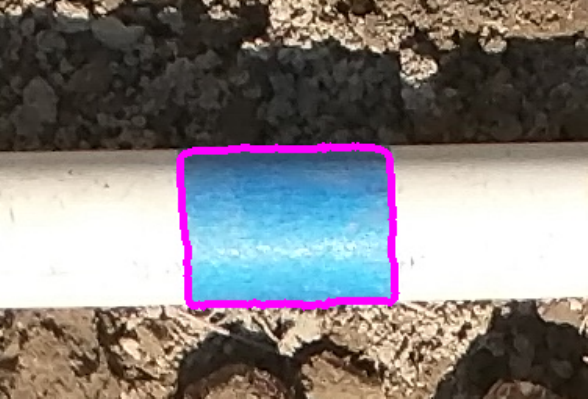}\label{fig:outline}}\cr
}
\caption{Plots were divided into 3 sections ((\subref{fig:plot_1}), (\subref{fig:plot_2}), and  (\subref{fig:plot_3})) with the PVC frame and each section was imaged. Images were captured about 1 meter above the soil, approximately perpendicular to the ground. Each section consisted of approximately 10-20 SPs. Blue and red tape were used for spatial calibration. An off-axis view of the entire plot is shown in (\subref{fig:offaxis}). In (\subref{fig:outline}), the magenta outline shows the tape's detection by color thresholding. }
\label{fig:plot_3_parts}

\end{figure}

Prior to data collection and harvest, vines were removed with a vine mower, and the SPs were dug up with a chain digger. The freshly dug SPs were then manually spread out over the beds. In total, 540 plot-level cellphone images of SPs were captured for the study using a Google Pixel 4A, with a pixel count of $4160\times3120$ pixels. Of these, 500 images consisted of plots divided into three sections, which was facilitated through the use of a polyvinyl chloride (PVC) plastic pipe "reference" frame. The frame was 10 ft long and contained 3 sections, each with a length of 3.3 ft. Figure \ref{fig:plot_3_parts} shows the frame's layout for one plot. For the remaining images, the entire plot and frame were imaged at an angle. Images were also captured at different times of day with varying cloud cover.

After digital images of the plots were taken, the SPs in each plot were harvested into agricultural containers and transported to the HCRS where they were scanned with a commercial optical sorter (Exeter Engineering, Exeter, CA) at the Horticultural Crops Research Station.
The calibrated sorter provided both RGB and near infrared (NIR) camera imagery from two orthogonal views of each SP to provide a comma separated value file containing each SPs length, width, and weight estimates. All metrics were also associated with each unique plot ID via the breeding program's existing barcode system. File names were used to pair cellphone images to plot IDs.

\begin{figure}[h]
    \centering
    \begin{subfigure}{0.757\textwidth}
    \centering
        \includegraphics[width=\textwidth]{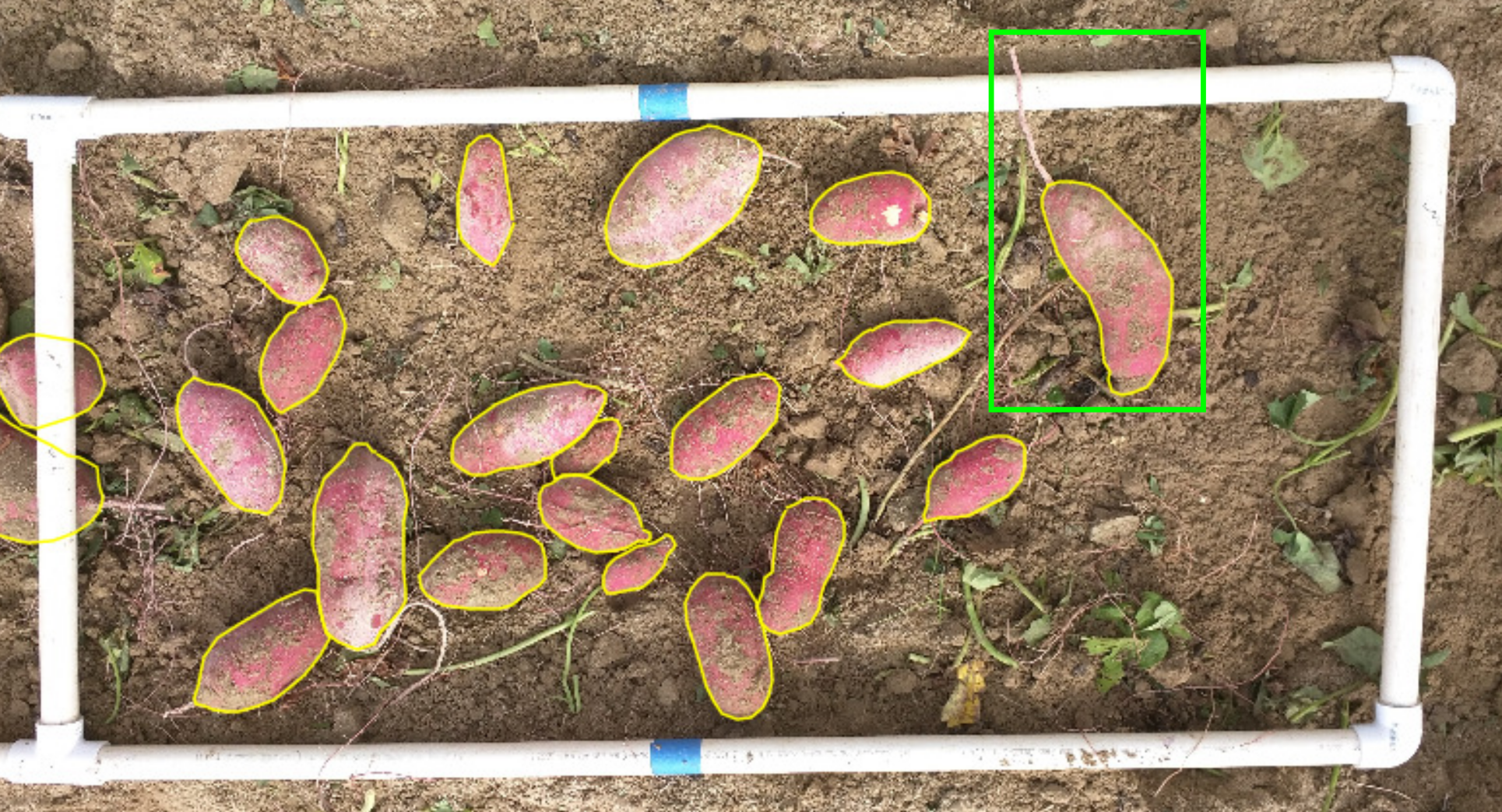}
        \caption{\label{fig:annotation_1}}
    \end{subfigure}
    \begin{subfigure}{0.232\textwidth}
    \centering
        \includegraphics[width=\textwidth]{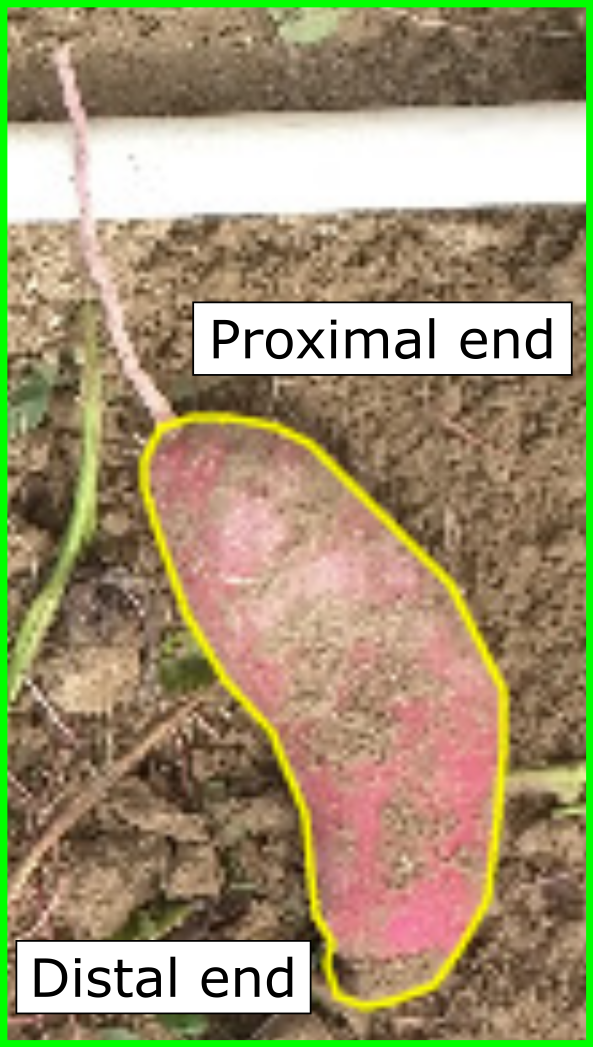}
        \caption{\label{fig:annotation_2}}
    \end{subfigure}

    \caption{\label{fig:annotation_all}In the example annotation (\subref{fig:annotation_1}), SPs are manually outlined in yellow using the VIA software.  Highlighted in (\subref{fig:annotation_2}), annotations are rounded over at the proximal and distal ends, ignoring root stalks and other thin regions.}
\end{figure}

\paragraph{Image Dataset Labeling}
SPs in the cellphone images were manually annotated, capturing outlines of each SP using the VGG Image Annotator (VIA) software \cite{VIA} as illustrated in Figure \ref{fig:annotation_1}. Because this software could be used directly in a web browser without installation, it was readily deployed to several annotators. In total, approximately 12,000 instances of SP roots across 540 images were annotated for this model. Annotators were instructed to outline only what they could see. For instance, SPs that were overlapping or occluded by other cull material (dirt, vines, etc.) were not inferred by the outlines. Similarly, SPs that may have been partially obscured by an overlapping SP was not inferred. Finally, hanging roots were also excluded from the masking as per Figure \ref{fig:annotation_all}(\subref{fig:annotation_2}).

\paragraph{Training Parameters}
A total of 540 images were used to build the model, with 60\% used for training, 20\% for validation, and 20\% for testing. The test set was not used in building the model. A COCO-based baseline was chosen with a backbone consisting of a 50-layer residual neural network (ResNet-50) and a feature pyramid network trained at a 3X learning rate. Notable is that Resnet-50 was selected over Resnet-101 to minimize the model's size, training time, and inference time. Of the 50-layer models, this backbone combination yielded the best performance concerning mask and box average precision (AP) when evaluated on the COCO dataset \cite{mscoco}.  

\paragraph{Spatial Calibration}
Attached to each section of the PVC plastic pipe grid was a strip of blue or red tape. This was used as a spatial reference to calibrate each image and enabled the number of centimeters per pixel to be ascertained in each picture. Color thresholding was used to differentiate the tape from the soil. For off-NADIR images, the tape closer to the camera was used. Figure \ref{fig:outline} shows the tape's detection. From this region, the minimum bounding rectangle was computed, providing the region's width and height. From the tape's known dimensions, a conversion factor, in units of pixels per centimeter, was calculated for each image.  Pixel counts, calculated from the instance segmentation results (masks), were converted to absolute distance units using this conversion factor.

\subsubsection{Model Validation}
\label{Model Validation}
\paragraph{Commercial Optical Sorter Baseline Performance Assessment}
To establish a baseline performance for the optical sorter, ground-truth measurements were made. Lengths, diameters, and weights of 240 roots were measured both manually and using the sorter. A regression analysis for each metric was performed to quantify estimation error.  

\paragraph{Plot-level Validation}
\label{plot_level}
To validate the trained model in Section \ref{Model Training}, diameter, length, and weight distributions were compared to the optical sorter's estimates, using images in the test set.   For this section and the remainder of the paper, weight estimates from the cellphone model use the ellipsoid estimate from Eq. \ref{eq:ellipse_volume}. 

\paragraph{One-to-one Validation}
In section \ref{plot_level}, measurements of the grid images from the optical sorter had a plot-level resolution.  To individually assess the performance of the cellphone model to an optical sorter, a second experiment was conducted that enabled resolution to individual SPs. To pair an individual SPs' measurements between the cellphone data and the commercial optical sorter, 110 SPs were labeled with a unique identification number. Each SP was then run sequentially through the sorter.  A patch of soil was then tilled up to use as a relevant background for imaging. Approximately 20 SPs were randomly selected and placed in the patch, label-side-down, such that the label did not interfere with the segmentation. Similar to the imaging protocol, a PVC pipe with blue tape was used to provide a spatial calibration.  After imaging the SPs with the cellphone, they were rotated so that the labels were visible and readable in the cellphone's image. Another image was taken for this arrangement, such that each SP's mask in the first image could be associated with the unique ID and subsequently matched to the data taken using the optical sorter.  In total, 30 pairs of images were collected using this procedure.

\section{Results}

\subsubsection{3D-to-2D Monte Carlo Simulations}\label{res:monte-carlo}

\begin{figure}[hbtp]
    \centering
    \includegraphics[width=\textwidth]{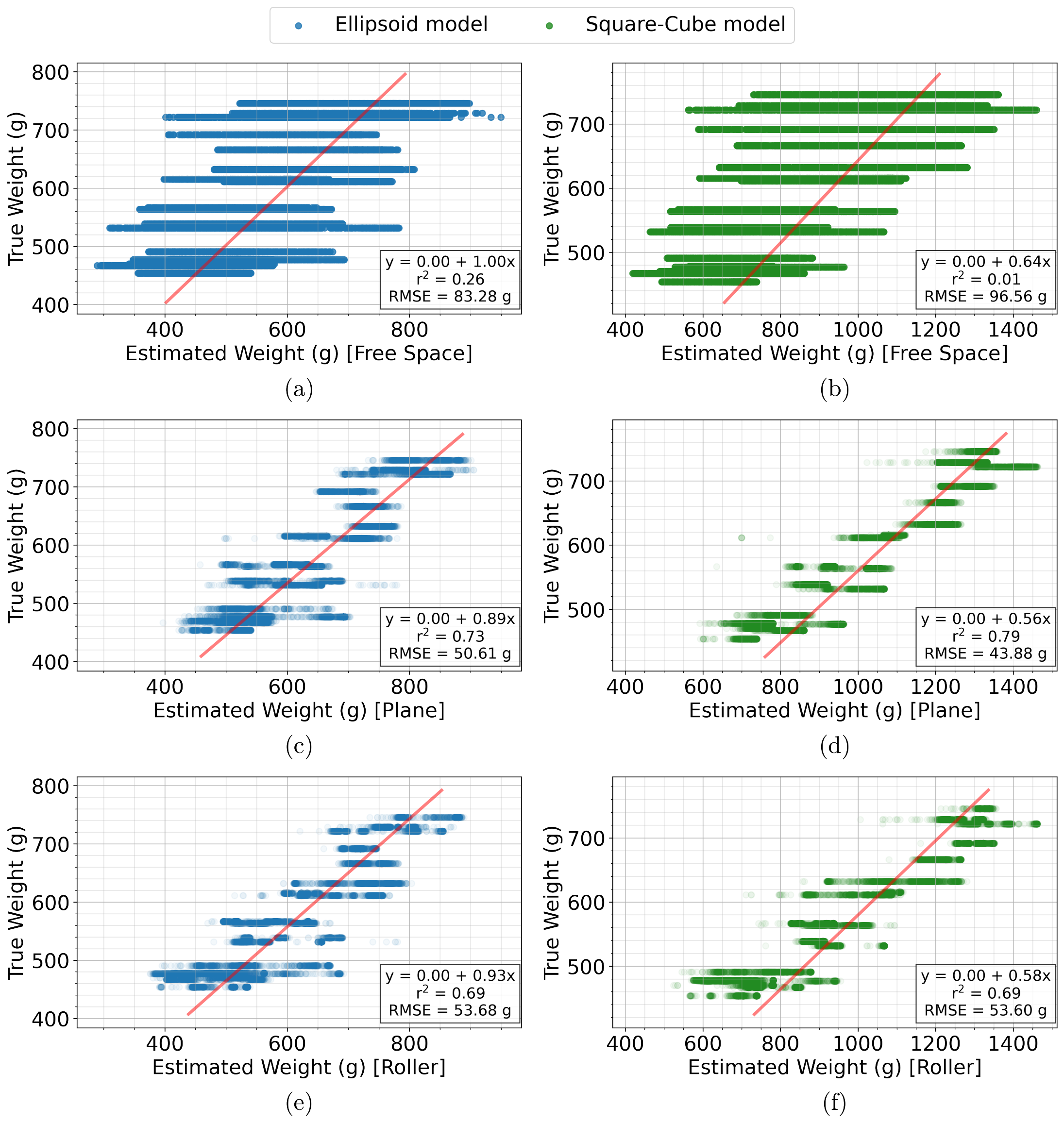}
    \caption{Results for the Monte Carlo simulations in free space, plane, and roller are shown in the top, middle, and bottom rows, respectively. Left and right plots show estimates from the ellipsoid and square-cube model, respectively. Each point represents a single root. A regression line was fit to each simulation, fixing the y-intercept to zero. r$^2$ and unbiased RMSE values are displayed.}
    \label{fig:sim_3_parts}
\end{figure}

Results for the simulation are shown in Figure \ref{fig:sim_3_parts}.  Predicted weights using all possible 2D projections were compared to the true weights of the roots using the ellipsoid and square-cube models in Figure \ref{fig:sim_3_parts} (a) and (b), respectively. Generally, performance was poor due to the variance in projected area when viewing a major-minor axis projection versus a minor-minor axis projection. By constraining the SPs, on either a plane (Figure \ref{fig:sim_3_parts} (c) and (d)) or with rollers (Figure \ref{fig:sim_3_parts} (e) and (f)), the set of possible projections is reduced, and r$^2$ and RMSE greatly improve.  The linear model under-predicted smaller root weights and over-predicted larger root weights. From this dataset, its uncertain whether this trend will continue for roots weighing less than 400 g. Visually, between the ellipsoid model and square-cube model, there does not appear to be a significant difference.   

\subsubsection{Plot-Level Phenotyping Performance}\label{plot_level}

\begin{figure}[H]
    \centering
    \includegraphics[width=1\textwidth]{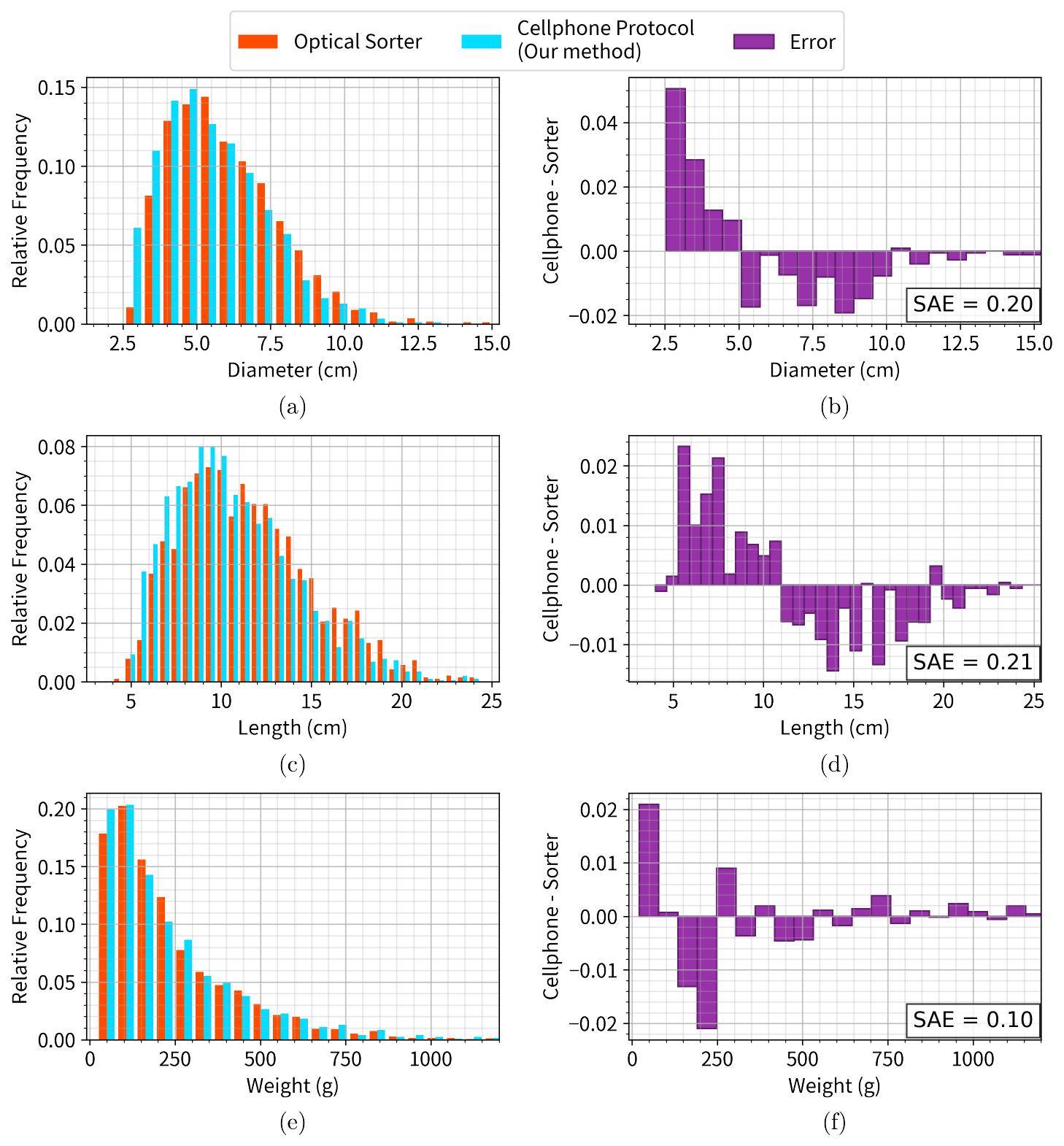}
    \caption{From the 2019 and 2020 dataset, distributions are shown for length, diameter, and weight counts as well as the corresponding absolute errors, as measured by both the cellphone method and the optical sorter.  Bin widths of 0.635 cm for the length and width and 56.70 g for the weight were used. The sum of the absolute error (SAE) is shown for each error plot.}
    \label{fig:plot_level_res}
\end{figure}

Distributions of the diameter, weight, and length estimates from the optical sorter and the cellphone protocol are depicted in Figure \ref{fig:plot_level_res}. Because the optical sorter is unable to measure roots under a certain size, roots smaller than 2.54 cm in diameter or 5.08 cm in length were removed from the optical sorter's dataset and the dataset collected with the cellphone protocol.  Figure \ref{fig:plot_level_res} (a) depicts the relative frequency of roots with a given diameter for both the optical sorter and the cellphone protocol, alongside the error in the frequency for each bin in Figure \ref{fig:plot_level_res} (b). Similar representations are depicted in Figure \ref{fig:plot_level_res} (c) and Figure \ref{fig:plot_level_res} (e) for length and weight with error frequencies depicted in Figure \ref{fig:plot_level_res} (d) and Figure \ref{fig:plot_level_res} (f), respectively. Generally, the error was the largest for the smallest roots across all metrics.

\begin{figure}[H]
    \centering
    \includegraphics[width=0.5\textwidth]{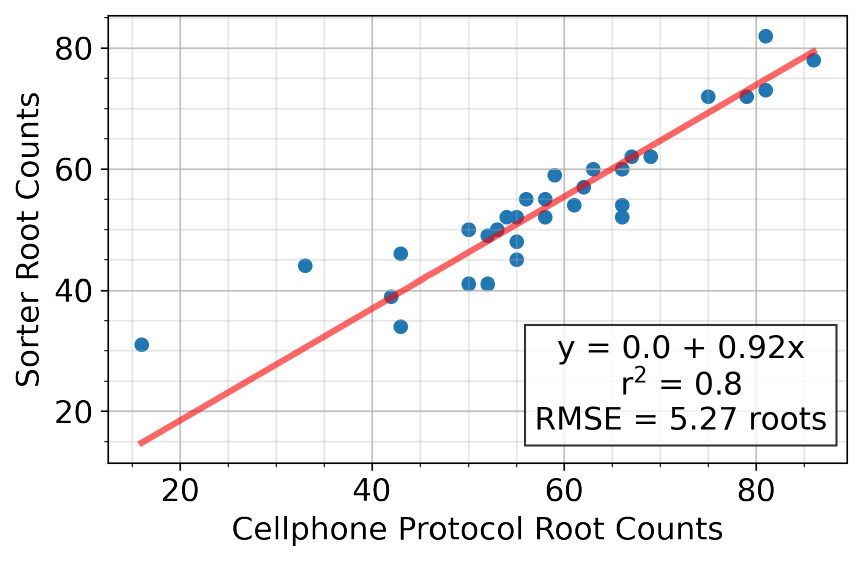}
    \caption{Root counts are shown for each of the 33 plots. A regression line is fit to each simulation, fixing the y-intercept to zero. r$^2$ and unbiased RMSE values are displayed. Each point represents a single plot.}
    \label{fig:counts}
\end{figure}

A comparison of counts between the sorter and cellphone model are shown in Figure \ref{fig:counts}. Data-points where the number of images did not match the number of sub-plots were removed. The scatterplot show a high correlation for counts between the models.

\subsubsection{One-to-One Phenotyping Performance}\label{res:one-one}

\begin{figure}[H]
    \centering
    \includegraphics[width=1\textwidth]{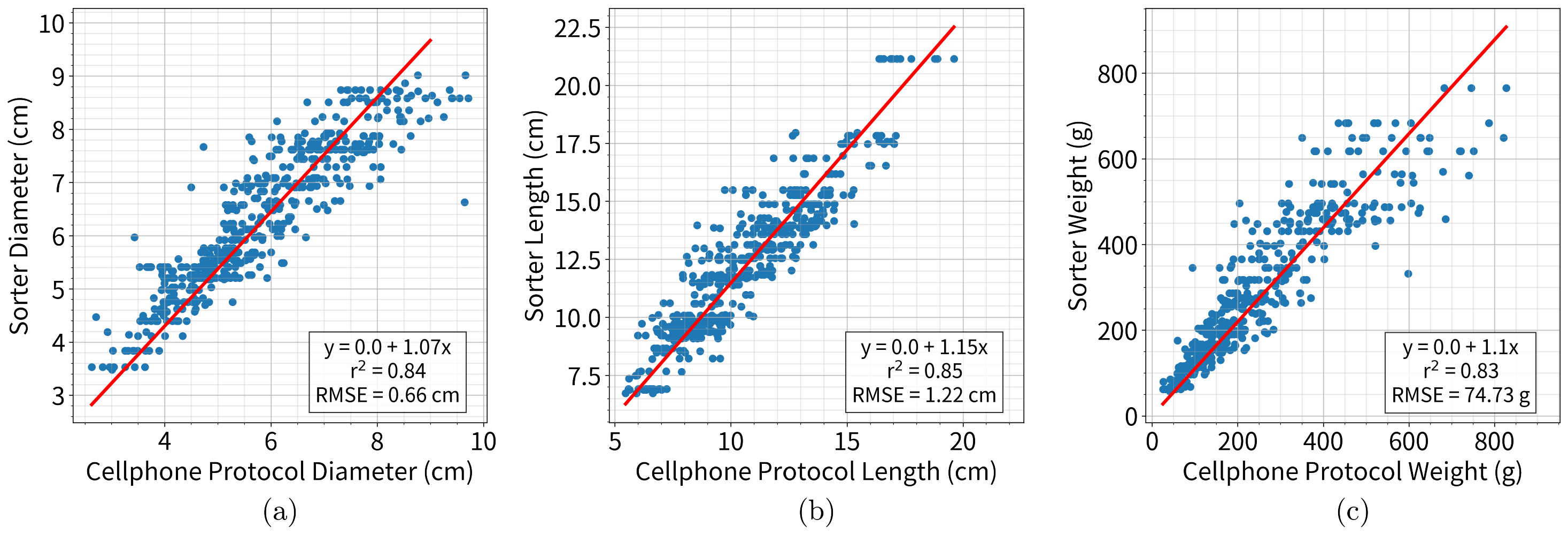}
    \caption{Results for the one-to-one simulations for the lengths, diameters, and weights are shown. A regression line is fit to each simulation, fixing the y-intercept to zero. r$^2$ and unbiased RMSE values are displayed. Each point represents a single view of a SP.}
    \label{fig:correlations}
\end{figure}

Results from the one-to-one study, displayed in Figure \ref{fig:correlations}, exhibit a high correlation between the aforementioned metrics estimated by the cellphone model and the optical sorter. Notably, the length RMSE is much larger than the diameter RMSE by a factor of 1.2.  This could be attributed to changes in perceived length due to the limited 2D view. Additionally, there is a bias in the model's length and width estimates in that the model underestimates these dimensions by about 8 to 15\%. The weight is also underestimated, on average, by about 8\%.

For the estimates of length and diameter, it appears that the variance is fairly consistent across sizes. However, with regards to weight, for larger SP, there is higher variance. Because this variance seems to increase with weight (heteroscedastic), it is difficult to accurately assess the regression.  Likely, the cause of the change in variance is due to abnormal shapes in the sample. From the Monte Carlo analysis, there was not a large change in variance across the 400-800 g range. However, in that test, fewer, more uniform root samples were simulated. It is possible that a more complex shape model would be able to further account for this variance.

\subsubsection{Commercial Optical Sorter Baseline}

\begin{figure}[H]
    \centering
    \includegraphics[width=1\textwidth]{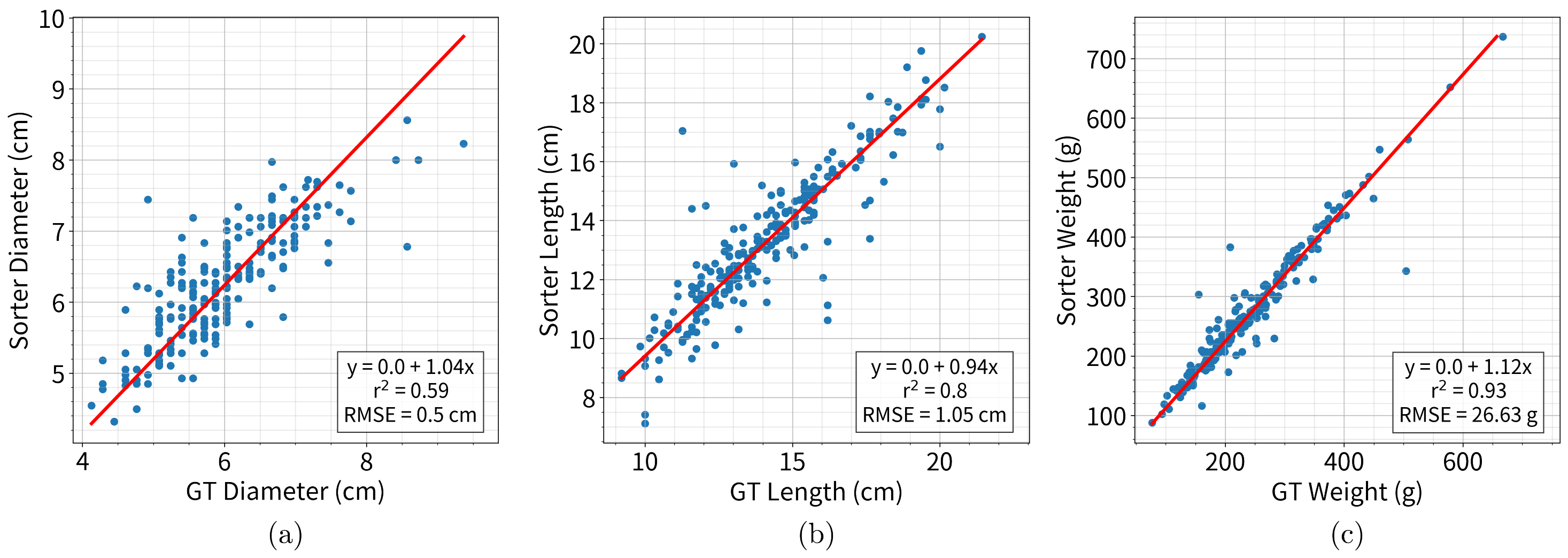}
    \caption{Results for the baseline for the lengths, diameters, and weights are shown here. r$^2$ and unbiased RMSE values are displayed. Each point represents a single SP.}
    \label{fig:gt_kylie}
\end{figure}

Comparisons between ground truth and sorter estimations are displayed in Figure \ref{fig:gt_kylie}.  The results from the manual measurements suggest that there is a strong correlation between the sorter's estimates for the length and weight (Figure \ref{fig:gt_kylie} (b) and (c), respectively), but a weaker correlation for diameter estimates (Figure \ref{fig:gt_kylie} (a)).  This makes sense because the optical sorter obtains two simultaneous views of the product, rather than one, to make predictions; thus, a lower weight estimation error is expected since the width dimension is more likely to change than the length when the SP is rotated by 90 degrees.

\section{Discussion}\label{discussion}
 From the plot-level results (section \ref{plot_level}), the distributions for the estimation metrics and the counts seem to agree between the cellphone model and the sorter as a whole. In this section, the anticipated error sources of the cellphone imaging protocol are quantified. Our results suggest that this model provides consistent estimates compared to those of the optical sorter.  Key error sources are enumerated to quantify the model's performance. While many of the individual error sources are inseparable, others are directly accessible from these results or theory.  
\subsubsection{Error Sources}

\begin{table}[h]
\centering
\caption{Error sources and estimated contributions}

        \begin{tabular}{cllcc}
            \hline
            \multicolumn{1}{c}{\textbf{Index}} & \multicolumn{1}{l}{\textbf{Error Metric}} & \multicolumn{1}{l}{\textbf{Category}} & \multicolumn{1}{l}{\textbf{Proportional Bias}} & \multicolumn{1}{l}{\textbf{RMS Error (g)}} \\
            \hline
            \multicolumn{1}{c}{\rownumber} & \multicolumn{1}{l}{Solid Angle} & \multicolumn{1}{l}{Imaging} & \multicolumn{1}{c}{0.99} & \multicolumn{1}{c}{{-}} \\ \hline
            \multicolumn{1}{c}{\rownumber} & \multicolumn{1}{l}{Lens Distortion} & \multicolumn{1}{l}{Imaging} & \multicolumn{1}{c}{0.95} & \multicolumn{1}{c}{-} \\ \hline
            \multicolumn{1}{c}{\rownumber} & \multicolumn{1}{l}{Lighting and Shadows} & \multicolumn{1}{l}{Masking} & \multirow{3}{*}{0.88}  & \multirow{3}{*}{54.98} \\ \cline{1-3}
            \multicolumn{1}{c}{\rownumber} & \multicolumn{1}{l}{Occluding Material} & \multicolumn{1}{l}{Masking} &  \\\cline{1-3}
            \multicolumn{1}{c}{\rownumber} & \multicolumn{1}{l}{Root Density} & \multicolumn{1}{l}{Modeling} & \\\hline
            \multicolumn{1}{c}{\rownumber} & \multicolumn{1}{l}{Root Shape} & \multicolumn{1}{l}{Modeling} & \multirow{2}{*}{1.12} & \multirow{2}{*}{50.61} \\\cline{1-3}
            \multicolumn{1}{c}{\rownumber} & \multicolumn{1}{l}{Volumetric Approximation} & \multicolumn{1}{l}{Modeling} & \\\hline 
            & & \multicolumn{1}{l}{Total:} & \multicolumn{1}{c}{0.93} & \multicolumn{1}{c}{74.73} \\\cline{3-5}

        \end{tabular}
    \label{tab:error}
\end{table}

 To quantify the model's performance, key bias and noise error sources are enumerated in Table \ref{tab:error}. Since yield is a key interest to the stakeholders, the values presented here depict error for weight estimation in prior experiments, although similar error breakdowns for length and diameter estimates can be achieved.  From Figure \ref{fig:sim_3_parts} (d) and (f), the square-cube model's performance suffered significantly from proportional bias, while the ellipsoidal model performed significantly better. For this reason, results from ellipsoidal model are referenced here for discussion on error. The one-to-one experiment in section \ref{res:one-one} provide an approximation of the total error in weight estimation while the Monte-Carlo experiment in section \ref{res:monte-carlo} provided an approximation of the errors induced by the root shape and geometric approximation. Errors are divided into proportional bias terms (proportional overestimation or underestimation) and an RMS error. RMS error is assumed to be uncorrelated such that each term is added in quadrature.

\subsubsection{Imaging Error Sources}
Imaging error sources are defined as those that distort the apparent root's size in an image from its true size.  In the Monte-Carlo simulations, 2D projections of the root were used to form an image assuming parallel rays of light.  In reality, only a proportion of the surface area (solid angle) can be seen, since the root is not flat. For a sphere, the ratio $S$ of the perceived projected area to the maximum projected area can be derived by the sphere's radius $R$ and distance to the sphere's front surface $D$ as:   

\begin{equation}
S = \frac{(\frac{D}{R} + 1)^2 - 1}{(\frac{D}{R} + 1)^2}
\end{equation}

For an imaging distance $D = 1$ m and root radius $R = 3$  cm, $S = 0.999$.  That is, less than 0.1\% of the area is lost.  Although there will be variation in magnitude based on the root's shape and orientation, noise from this error source is assumed to be negligible.  

Lens distortion due to the imaging optics also contributes to imaging errors. Images can appear warped, with parallel lines appearing curved.  When measuring roots in the image, this tends to add greater error for roots near the field of view's (FOV) edges compared to roots located at the center. For example, for 5\% barrel distortion, roots at the image's edge will appear approximately 5\% smaller.  
Since distortion is negligible at the center of the FOV, and lenses are generally well-corrected, this error source was mitigated by using the cell phone's zoom lenses (not wide-angle lenses) to minimize distortion. Similar to the solid angle, minimally significant error was assumed from the distortion, which would be dependent on the lens design and camera.

\subsubsection{Masking Error Sources}
Error contributions from external factors, such as shadows and occlusions, can also occur which may reduce the masks' quality. Assumptions include accurate training annotation outlines and model masks that do not significantly deviate from the outlines. To estimate the contribution from these error sources, other error sources (imaging, modeling) were first quantified before a quadrature subtraction of these contributions from the total error.  The total error was approximated using results from the one-to-one experiment in Figure \ref{fig:correlations} (c). 
In Table \ref{tab:error}, errors due to root density are included along with the masking terms, as root density was not simulated in the Monte Carlo simulations. Overall, the noise contribution (54.98 g) from these terms appears to be on par with the modeling terms. Even though completely eliminating the masking error sources could be challenging, these results imply that performance could be significantly improved by reducing the modeling errors with a more complex model incorporating density or shape terms. The bias term here (0.88) was calculated by de-biasing the regression from the one-to-one results and dividing the other bias terms. This term implies an underestimation of the weight by 12\% due to these factors. As noted in the subsequent section, overestimation from the modeling terms seems to offset this bias.

\subsubsection{Modeling Error Sources}
\label{Monte_carlo}
From the Monte Carlo simulations, described in section \ref{Monte Carlo}, variability in the modeled weight versus projection angle (or area) is simulated, independent of the imaging and environmental errors. In this section, the error contribution from the experiment were quantified due to the variation in the roots' shapes and orientations.   Generally, the Monte Carlo simulations showed reduced variability when the SPs' 2D projections were physically constrained, yielding an increased correlation with root weight.  Using an ellipsoid model, an RMSE of 50.61 g, shown in Figure \ref{fig:sim_3_parts} for weight estimates, was achieved, assuming the roots were constrained to a flat surface. However, even with these constraints, both volumetric models tended to overestimate the weight. In Table \ref{tab:error}, the ellipsoid model's proportional bias was calculated to be 1.12. Such proportional bias can be removed by dividing all of the model's estimated outputs by this proportionality term. Furthermore, the ellipsoid model's RMS error and proportional bias were not significantly different for both the plane and roller constraints.  This was due, in part, to the roots' circular symmetry. Conversely, the square-cube model was more likely to underestimate smaller roots. Equation \ref{eq:volume} makes several assumptions about the 3D shape, such as an ellipsoidal structure and circular cross sections. For a misshapen root with an irregular cross-section, its thickness (in the direction parallel to a camera) differs from its diameter. Because a root is more likely to rest on its wider dimensions, in a 2D image, its thickness is overestimated, and consequently, its weight is also overestimated.

\subsubsection{Plot-Level Phenotyping Performance}
Despite the errors present within the in-situ study, the distributions of length, width, and weight between the optical sorter and the cellphone model are in agreement. A Chi-square test for goodness of fit indicates that the proportions of diameters, as estimated by the masking method, were consistent to the proportions measured by the optical sorter ($\chi^2 = 0.28, df = 12, p = 1.0$). Similarly, the goodness of fit tests for lengths ($\chi^2 = 0.09, df = 22, p = 1.0$) and weight ($\chi^2 = 0.15, df = 21, p = 1.0$) distributions suggest that there is no significant difference between this method's estimates and the optical sorter's estimates. As the Chi-square test requires a reasonably large sample size, bin pairs where either bin had fewer than 10 counts were discarded. In total, this was about 10\% of the total samples. The model tends to overestimate smaller roots in all 3 parameters compared to the sorter. One cause of this may be due to how storage roots increase in size. In earlier stages, the storage roots first develop in length along the major axis. At later stages bulking is more apparent, where growth will occur along the minor axis. Across these stages, as the aspect ratio decreases, the ellipsoid approximation shifts from overestimating to underestimating the root's volume. Across the test sets, qualitatively, the segmentation appears to perform fairly well against harsh shadows and soil clumps.

\section{Conclusion}
  Because sweetpotatoes can vary greatly in skin color and shape across clones, it is challenging to utilize traditional computer vision techniques to produce accurate detections. A Mask R-CNN model was developed to perform instance segmentation on sweetpotatoes that were dug up and singulated. This model accurately counted roots (r$^2$ = 0.8, RMSE = 5.27 roots per plot) and estimated length, width, and weight from cellphone images (RMSE = 0.66 cm, 1.22 cm, and 74.73 g) as compared to a commercial optical sorter.  As opposed to an optical sorter, this approach could be readily deployed in a portable, modular system.  By capturing images using a handheld device (e.g. cellphone), unmanned aerial vehicle (UAV), or other means, yield estimates for entire fields would be possible without a dedicated sorter. With the trend towards big data in agriculture and integration of sensor technologies in precision farming, the ability to monitor yield and growth would provide both breeders and growers vital feedback. For distribution, with regards to warehousing processes, efficient order picking and packing to fulfill customer orders is essential to minimizing logistical costs.  Because SP quality (e.g. shape, size, defects) may vary greatly, assessing quality at intermediate stages of the supply chain during packing or even storage would enable better planning.  For deployment in such areas, future research is needed to investigate the challenges of close packing (occlusion) of roots on the accuracy of segmentation.   
  
\section*{Acknowledgments}

\subsection*{General} 
We would like to thank members of the Sweetpotato Breeding and Genetics Program: Mark Watson, Simon Fraher, and Modesta Abugu for imaging SPs used in our model and providing feedback for our studies.  We would also like to thank staff at the HCRS in Clinton for facilitating the rapid data collection of SPs.

\subsection*{Author Contributions} 
H. Nguyen and M. Kudenov conceived the model pipeline, designed experiments, and wrote the manuscript. S. Gyurek annotated images used for model training. R. Mierop and K. Pecota led the data collection process for both the cellphone imaging protocol and optical sorter. K. LaGamba and M. Boyette conducted baseline measurements for the optical sorter used in this paper. G. Yencho and C. WIlliams provided direction towards relevant research objectives and experimental design.   

\subsection*{Funding}
This work was supported by the North Carolina State University GRIP4PSI Program [grant number 573000] and the National Science Foundation [grant number 1809753].

\subsection*{Conflicts of Interest}
The authors declare that there is no conflict of interest regarding the publication of this article.

\subsection*{Data Availability}
The source code for the Blender simulation and segmentation model are available upon reasonable request.

\printbibliography

\end{document}